\documentclass[prb,twocolumn,amsmath,amssymb,showpacs]{revtex4}
\pdfoutput=1
\usepackage{graphicx}
\usepackage{dcolumn}
\usepackage{bm}

\begin{document}

\title{Lattice expansion from isotope substitution in the Iron based superconductors}
\author{Oleg Kim,$^1$ and Mats Granath$^2$}
\affiliation{$^1$Department of Applied Physics, Chalmers University of Technology,
SE-41296 Gothenburg, Sweden}
\affiliation{$^2$Department of Physics, University of Gothenburg,
SE-41296 Gothenburg, Sweden}

\date{\today}

\begin{abstract}
We estimate the changes in lattice parameters due to iron isotope substitution in the iron-based high temperature superconductors using 
a variational calculation based on the anharmonicity of the Fe-As(Se) bond. 
For BaFe$_2$As$_2$ and $^{57}$Fe to $^{54}$Fe substitution we find a c-axis expansion of $1\cdot 10^{-3}${\AA} and 
in-plane a-axis contraction of $1\cdot 10^{-5}${\AA} at 250K in good agreement with experimental values on (Ba,K)Fe$_2$As$_2$ 
for which a negative isotope exponent $\alpha=-\frac{d\ln T_c}{d\ln m}<0$ was found [P.M. Shirage {\em et al.}, Phys. Rev. Lett. {\bf 103}, 257003 (2009)]. For FeSe in contrast we find an expansion for both a and c lattice parameters. 
We discuss the relevance of these isotope induced changes of lattice parameters to the isotope effect in light of the well established sensitivity 
to the Fe-As(Se) coordination.
\end{abstract}


\maketitle

There is evidence to suggest that the iron-based superconductors\cite{Kamihara} are unconventional in the sense that superconductivity is not caused by electron-phonon interactions.\cite{review} Pointing in this direction is the close proximity, 
or even coexistence, of magnetic order or strong magnetic fluctuations\cite{Zhao} as well as the results from density functional theory (DFT) of 
mundane electron-phonon interactions.\cite{Boeri}

With this in mind, the observations of significant effects on T$_c$ with iron isotope substitution is remarkable. In the first study of the isotope effect by 
Liu {\em et al.} an isotope exponent $\alpha=-\frac{d\ln T_c}{d\ln m}\approx 0.35$ for iron substitution in K doped BaFe$_2$As$_2$ and F doped SmFeAsO was found\cite{Liu} suggestive of a more or less conventional electron-phonon pairing mechanism.  
In subsequent work by Shirage {\em et al.} on nominally the same (K,Ba)Fe$_2$Fe$_2$ a startling negative isotope exponent  $\alpha\approx -0.18$ 
was found.\cite{Shirage_PRL} Such a negative isotope effect is highly unconventional while at the same time the large magnitude clearly indicate a non-negligible influence of the lattice on superconductivity. The few other studies of the isotope effect on T$_c$ in these materials have ranged from tiny, 
$\alpha\approx -0.02$, in optimally doped SmFeAsO$_{1-y}$\cite{Shirage_2010} to large positive, $\alpha\approx 0.81$, in FeSe$_{1-x}$.\cite{Khasanov_isotope}  It is of course possible given the still early stages of the study of these materials, that not all of these results represent genuine isotope effects. 

At the same time, the underdoped cuprate superconductors, which are the prototypical unconventional superconductors, also display large and doping dependent isotope exponents. \cite{isotope_cupr} Particularly intriguing is the strong enhancement around 1/8'th doping in La$_{2-x}$Sr$_x$CuO$_4$ where 
charge and spin density wave order which also couples to the lattice is known to influence superconductivity. Optimally doped cuprates, with the maximum transition temperatures, on the other hand show little or no isotope effect, a possible reason why this phenomenon has been less studied than many other exotic phenomena in these materials.

It is clearly an important but apparently largely unresolved issue what is the cause of isotope effects in these high temperature superconductors. 
One clue of the rich interplay between lattice and electrons is the  
tetragonal to orthorhombic structural phase transitions and the accompanying electronic orders that are found in both types of systems.\cite{Shen_anisotropy,Vojta}
In this work we consider one particular aspect of isotope substitution in the iron based superconductors, namely the effect on lattice parameters that follows from anharmonicity in any real material. An important question is to compare these calculations to experimentally measured isotope induced changes in lattice parameters.   

In previous work\cite{Granath} a B$_{1g}$, Fe dominated phonon (energy near 20meV)
was studied by Raman spectroscopy and the temperature dependence of the energy and lifetime was found to be well reproduced within a calculation of  
the phonon self energy due to phonon-phonon interactions induced by assuming a cubic anharmonic
correction to the Fe-As bond potential. A rough estimate of the isotope induced changes of lattice parameters were also made based on the expansion 
of an 
isolated Fe-As bond. Substitution with a lighter mass gives an expansion of the bond which was taken to imply a proportional expansion of both a- 
and c-axis lattice parameters. The calculation with an isolated bond may give an order of magnitude estimate but to get a more accurate description it is
important to place the anharmonic bond in the appropriate crystal environment of an harmonic lattice. This is done in the present work. The main result 
is that for realistic parameters the expansion of the Fe-As bond will occur predominantly in the softer c-axis, accompanied with either expansion or contraction of the 
in-plane bond lengths. For BaFe$_2$As$_2$ we find changes that are quantitatively in line with those measured in the isotope measurements by Shirage {\em et al.} ($\delta_c=3\pm1\cdot 10^{-3}${\AA} and $\delta_a=0\pm 1\cdot 10^{-3}${\AA}) which gives independent support that those experiments were indeed done on properly isotope substituted samples.\cite{Shirage_PRL,Liu_comment}

To explain the calculation in a transparent fashion we will start by considering the one-dimensional anharmonic chain. 
The Hamiltonian for the problem is
\begin{equation}
H=\sum_i\frac{p_i^2}{2m}+\frac{k}{2}(r_{j+1}-r_j-a)^2-\frac{g}{6}(r_{j+1}-r_j-a)^3
\end{equation}
with $p_j$ and $r_j$ the momentum and position operators of ion $j$ and $a$ the equilibrium interatomic distance for the harmonic problem $g=0$. The aim being to
calculate the change in equilibrium distance $\delta=<r_{j+1}-r_j-a>$ for the anharmonic problem $g\neq 0$. 
The anharmonic oscillator is classic problem which cannot be solved exactly but 
for small $g$ it can be estimated variationally\cite{Feynman} in terms of the Hamiltonian $H'=\sum_i\frac{p_i^2}{2m}+\frac{k}{2}(r_{j+1}-r_j-(a+\delta))^2$   
and minimizing the free energy with respect to the variational parameter $\delta$. Thus minimizing the thermal expectation value $<H-H'>_{H'}$, using  
$r_j=j(a+\delta)+x_j$, such that $H'$ is now the standard Harmonic chain expressed in terms of the deviations $x_j$ from equilibrium positions, and
expanding to second order in $g$ gives 
\begin{equation}
\label{var1D}
2<H-H'>_{H'}=\sum_j k\delta^2-g\delta<(x_{j+1}-x_j)^2>_{H'}
\end{equation}
Since $k>0$ this has a minimum given by 
$\delta=\frac{g}{2k}\frac{1}{N}\sum_q 2(1-\cos q)<x_qx_{-q}>$ where $x_q=(1/\sqrt{N})\sum_j e^{iqj}x_j$. (With $N$ the number of sites in the chain.) 
The expansion is thus clearly related to the mean square displacement of the atom, or in the zero temperature limit the zero point motion. 
For the infinite chain, solving the harmonic chain in the standard fashion, we find
\begin{equation}
\label{expand}
\delta=\frac{\hbar g}{\sqrt{m}k^{3/2}2^{3/2}}\int_{-\pi}^{\pi}\frac{dq}{2\pi}\sqrt{1-\cos q}\coth\frac{1}{2}\beta\hbar\omega_q
\end{equation}
with $\omega_q=\sqrt{2(1-\cos q)k/m}$. The actual isotope substitution induced expansion is thus given by the difference in anharmonic expansions for the two masses 
$m_0$ and $m_1$, 
\begin{equation}
\Delta_{iso}=\delta_{m_1}-\delta_{m_0}\,.
\end{equation}  
In the zero temperature limit Eq. \ref{expand} reduces to $\delta(T=0)=\frac{\hbar g\pi}{\sqrt{m}k^{3/2}}$, with the corresponding isotope 
shift $\Delta_{iso}(T=0)=\frac{\hbar g\pi}{k^{3/2}}(1/\sqrt{m_1}-1/\sqrt{m_0})$ showing that substitution with a lighter mass ($m_1<m_0$) gives an expansion (for $g>0$). Although the anharmonic contribution to the lattice spacing, i.e. the thermal expansion, Eq. \ref{expand}, grows monotonically with increasing temperature, the actual isotope shift $\Delta_{iso}(T)$ decreases with temperature from a maximum at $T=0$, as the mass enters both in the prefactor $1/\sqrt{m}$ and in the phonon dispersion $\omega_q$.   

\begin{figure}
\includegraphics[width=8cm]{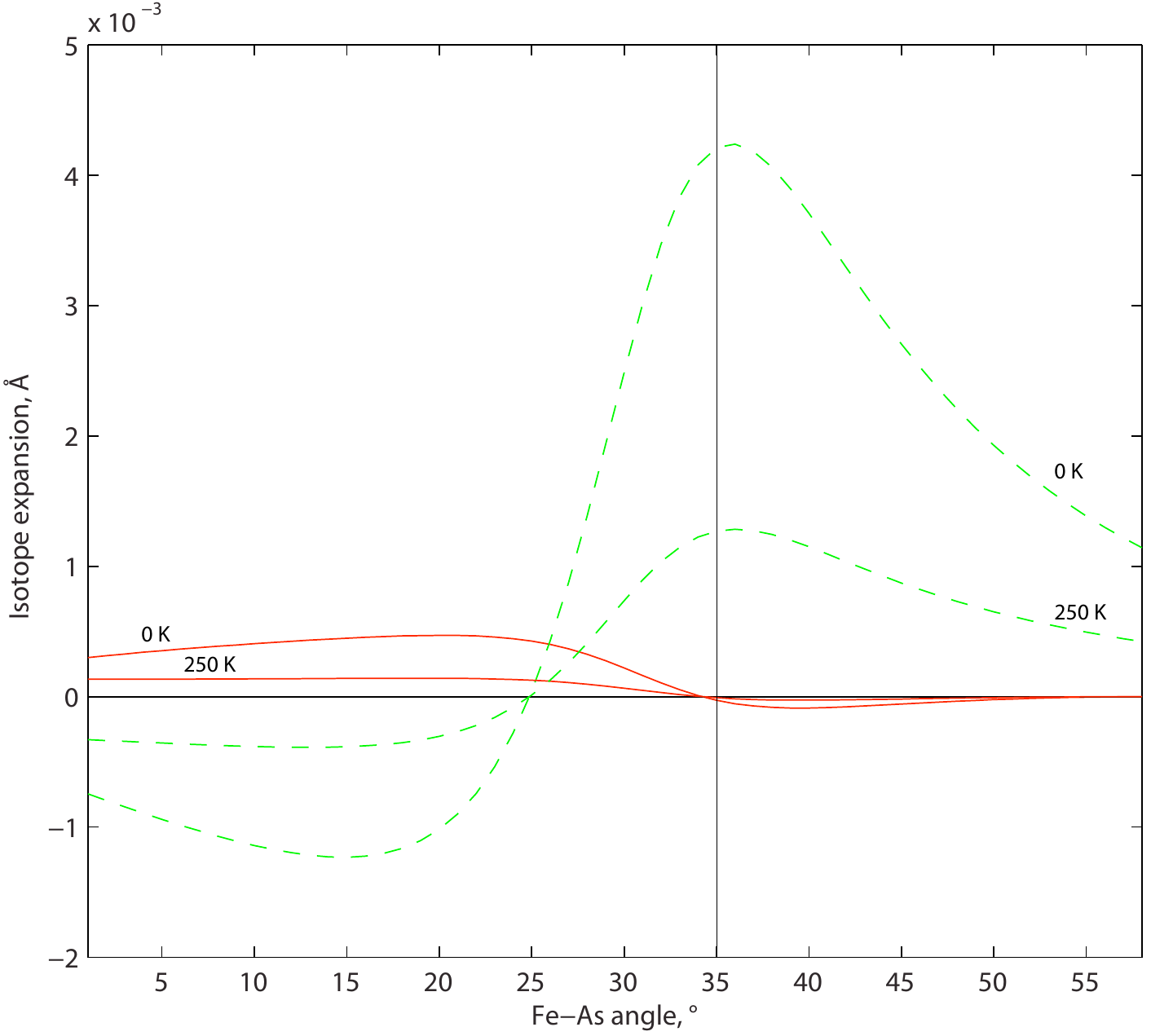}
\caption{\label{ba_fig}
Calculated expansion of lattice parameters of BaFe$_2$As$_2$ due to isotope replacement of $^{57}$Fe with $^{54}$Fe as a function of the elevation angle of As above the Fe plane, $\theta$. Expansion of c is dashed and a is solid. 
The bare values are $c\approx 13${\AA}, $a\approx 3.9${\AA} and $\theta$ $\approx 35^o$.\cite{Zhao}}
\end{figure}

\begin{figure}
\includegraphics[width=8cm]{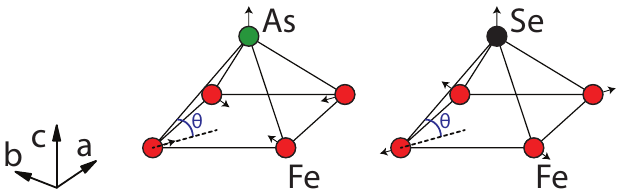}
\caption{\label{expansion_fig}
Sketch of calculated changes in lattice parameters from substitution with a lighter Fe isotope for BaFe$_2$As$_2$ (left) and FeSe (right).}
\end{figure}

For a three dimensional crystal the variational expression, Eq. \ref{var1D}, generalizes to 
\begin{eqnarray}
&&2\left\langle H-H' \right\rangle_{H'} =\\ 
&&\sum_{<ij>} k_{ij} (\hat{s}_{ij}\cdot\vec{\delta}_{ij})^2
- g_{ij}  (\hat{s}_{ij}\cdot\vec{\delta}_{ij}) <(\hat{s}_{ij}\cdot(\vec{x}_i - \vec{x}_j))^2>_{H'}\nonumber
\end{eqnarray}
where the sum runs over all bonds $i,j$ with harmonic (anharmonic) coupling constant $k_{ij}$ ($g_{ij}$), where $\hat{s}_{ij}$ is a unit vector in the bond direction, and where 
$\vec{x}_i$ is the displacement of ion $i$ with $\vec{\delta}_{ij}$ the variational expansion of the bond. The expectation value is then evaluated in terms of the full phonon spectrum of the crystal. 
We have performed the calculation for the BaFe$_2$As$_2$ crystal assuming harmonic bonds except for the nearest neighbor Fe-As bond
which is given a cubic anharmonicity. 
We assume a uniform expansion of the unit cell and consider only the tetragonal crystal structure. There are thus two variational parameters, the 
expansions $\delta_a$ and $\delta_c$ of the in-plane and out of plane lattice constants $a$ and $c$, respectively. Our harmonic model includes a minimal number of parameters needed to reproduce a realistic phonon spectrum 
with harmonic couplings as taken from the fit of the harmonic (Born-von Karman) model to experimental data of the phonon spectrum from 
inelastic X-ray scattering.\cite{Lee}
We use the nearest neighbor Fe-As bond ($k=$4.37 eV$/${\AA}$^2$), Fe-Fe ($k=$0.81 eV$/${\AA}$^2$), and 
Ba-As ($k=$0.50 eV$/${\AA}$^2$) and the equal height  As-As ($k=$0.25 eV$/${\AA}$^2$). The anharmonicity is $g=103$eV$/${\AA}$^3$ for the Fe-As bond.\cite{Granath} In Figure \ref{ba_fig} is shown the calculated changes of lattice parameters $\Delta_{iso,a}$ and $\Delta_{iso,b}$ from replacing $^{57}$Fe with $^{54}$Fe for two different temperatures. (We use $^{57}$Fe instead of the naturally more abundant $^{56}$Fe in order to compare to the experiments in Ref.\onlinecite{Shirage_PRL} where such samples were prepared.) As expected based on the solution to the 1D chain, the isotope shift decreases with temperature. The results are plotted as a function of the As elevation angle above the Fe plane $\theta$, see Fig. \ref{expansion_fig}, keeping the bare lattice parameters a and c fixed.  ($2\theta+\alpha=180^o$, with $\alpha$ the Fe-As-Fe tetrahedral angle.) The results are consistent with the expansion of the anharmonic Fe-As bond, which for small angles corresponds to expansion in the plane and for larger (physical) angles correspond to expansion in the c-direction. The actual magnitudes of expansions and contractions are related to the larger stiffness of the lattice in-plane than out-of-plane.

\begin{figure}
\includegraphics[width=8cm]{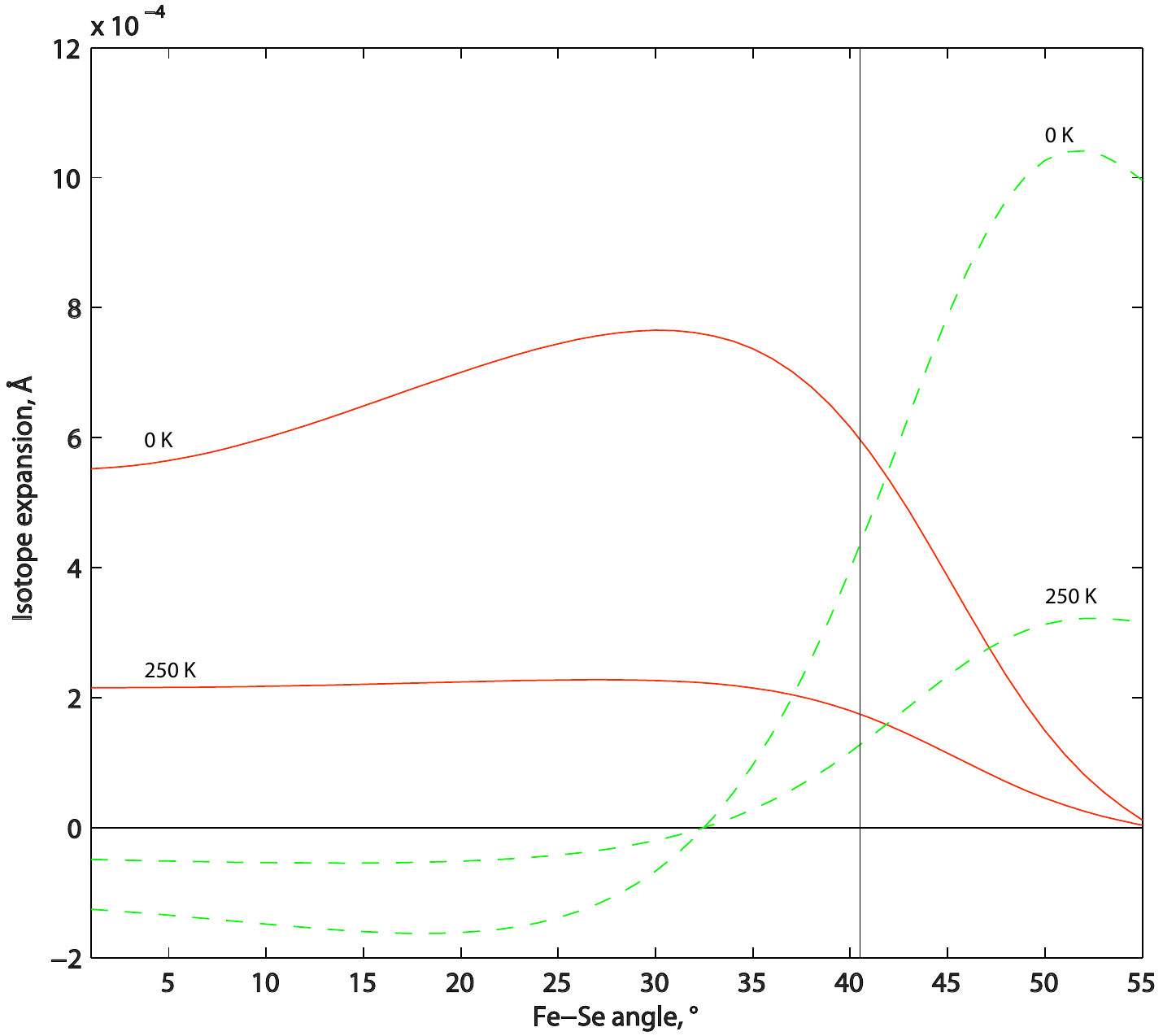}
\caption{\label{se_fig}
Calculated expansion of lattice parameters of FeSe due to isotope replacement of $^{57}$Fe with $^{54}$Fe as a function of the elevation angle of Se above the Fe plane, $\theta$. Expansion of c is dashed and a is solid. 
The bare values are $c\approx 5.5${\AA}, $a\approx 3.8${\AA} and $\theta$ $\approx 40.5^o$.\cite{Khasanov_isotope}}
\end{figure}

We have also considered FeSe with the expectation that the missing spacer layer would make the crystal less prone to expand in the c-direction. Here we were not able to find the relevant spring constants in the literature but made a rough fit to LDA calculated phonon spectra.\cite{Nakamura,fesebonds}  In contrast to the Ba122 crystal we did find an expansion in the plane of similar magnitude to the expansion in c as shown in Figure \ref{se_fig}.
Experimentally, a very large isotope effect $\alpha\approx 0.8$ was found in FeSe$_{1-x}$ together with a contraction in c and expansion in a of similar magnitudes $\sim 3-5\cdot 10^{-4}${\AA}.\cite{Khasanov_isotope} Given the uncertainty in our parameters for FeSe we tried making the interlayer couplings stronger but found that we had to make interlayer Fe-Se bond stronger than the corresponding intralayer bond in order to get a contraction in c together with an expansion in a. This may indicate that there are some problems with the isotope substituted samples used in Ref. \onlinecite{Khasanov_isotope}. In general, we have not found contractions in c for any reasonably physical set of parameters. However, possible sources of error in addition to the uncertainty in parameters could include only considering the undoped systems as well as the assumption of a uniform expansion of the unit cell.     

Clearly, isotope induced changes of lattice parameters are small, as estimated in this work and as seen experimentally\cite{Liu,Shirage_PRL,Shirage_2010,Khasanov_isotope}, at most of the order of a few times 
$10^{-3}${\AA} or of the order of a 0.1\% change. Based on such small changes of the crystal parameters it is difficult to see how to generate the 
large isotope shifts on the superconducting transition temperature in models for non-phonon mediated superconductivity.\cite{Granath} However, it has been argued in the past as well as recently that small changes in lattice parameters can indeed reproduce large isotope exponents\cite{Kivelson_isotope,Phillips}.   
In addition, a characteristic feature of the iron based superconductors is that they are extremely sensitive to small changes in the crystal structure such as the anion (As,Se,Te) height above the Fe plane.\cite{Kuroki} 
It was suggested by Khasanov {\em et al.} that the varying isotope exponents could be reconciled by having two contributions, one from 
the standard isotope shift due to Debye frequency shift and a separate contribution from changes in the anion height where for different systems isotope replacements would increase or decrease the height.\cite{Khasanov2010}
Although an interesting suggestion, our calculations do not support this picture, we always find a c-axis expansion for the lighter isotope for realistic parameters. 

We believe these calculations can provide a benchmark for experiments on isotope substituted samples by measuring the changes in lattice  parameters. If and how these small changes are enough to cause large isotope shifts of T$_c$ or whether genuine effects of phonons on the pairing need to be considered remains to be explored further.   

Finally we would also point out how the present calculations complement calculations of lattice parameters using density functional theory. 
DFT calculations cannot directly address the effects of isotope substitution 
as the atomic mass does not enter a calculation for the ground state energy of the electronic charge distribution.
Harmonic coupling constants and the corresponding phonon spectrum are of course readily available and needed as input to our calculation. Although the strength of anharmonicity of the bonds would also be possible to estimate we have not found any such studies on these materials, and have instead relied on an independent estimate of anharmonicity based on a many body calculations of the phonon self-energy and the corresponding temperature dependence of the energy. It is however possible to estimate the isotope induced expansion from DFT by including the zero point phonon energy in the total energy.\cite{Schimka} Although significantly more demanding than the present study such calculations would provide an interesting comparison. Another natural continuation of the present work on the isotope induced expansion is to explore the related thermal expansion\cite{Grimvall} which may also provide an independent experimental probe of the isotope substituted systems.


We acknowledge valuable discussions with S. \"Ostlund, G. Wahnstr\"om and H. Strand. 
The work was supported by the Swedish Research Council (grant no. 2008-4242).


\begin{thebibliography}{10}
\bibitem{Liu_comment} Measured changes of the c lattice parameter of the samples in Ref. \onlinecite{Liu} were all within the error bars.
\bibitem{fesebonds} We have used the intraplanar bonds Fe-Se ($k=$3.5 eV$/${\AA}$^2$), Fe-Fe ($k=$0.7 eV$/${\AA}$^2$), and 
and equal height Se-Se ($k=$0.21 eV$/${\AA}$^2$) and interplanar Se-Se and Fe-Se of equal magnitude ($k=$0.39 eV$/${\AA}$^2$).
\end{thebibliography}
\end{document}